\documentclass[usenatbib]{mn2e}

\usepackage{graphicx}
\usepackage{lscape}
\usepackage{txfonts}
\usepackage{url}

\title[Satellites around massive galaxies since $z \sim 2$]{Satellites around massive galaxies 
since $z \sim 2$}

\author[E. M\'armol-Queralt\'o et al.]{E. M\'armol-Queralt\'o$^{1,2}$\thanks{E-mail:
emq@iac.es}, I. Trujillo$^{1,2}$, P.G. P\'erez-Gonz\'alez$^{3,4}$, J. Varela$^{1,2,5}$ and  G. 
Barro$^{6}$\\
$^{1}$Instituto de Astrof\'{\i}sica de Canarias, c/ V\'{\i}a L\'actea s/n, 
E-38205, La Laguna, Tenerife, Spain\\
$^{2}$Departamento de Astrof\'{\i}sica, Universidad de La Laguna, E-38205, La 
Laguna, Tenerife, Spain\\
$^{3}$Departamento de Astrof\'{\i}sica, Facultad de CC. F\'{\i}sicas, Universidad Complutense de Madrid,
E-28040, Spain\\
$^{4}$Associated Astronomer at Steward Observatory, The University of Arizona\\
$^{5}$Centro de Estudios de F\'{i}sica del Cosmos de Arag\'on (CEFCA), Plaza San Juan, 1, Planta-2, E44001-Teruel, Spain\\
$^{6}$UCO/Lick Observatory, University of California, Santa Cruz, CA 95064}
\begin{document}

\date{}

\pagerange{\pageref{firstpage}--\pageref{lastpage}} \pubyear{2011}

\maketitle

\label{firstpage}

\begin{abstract}

Accretion of minor satellites has been postulated as the most likely mechanism
to explain the significant size evolution of the massive galaxies over cosmic
time. Using a sample of 629 massive ($M_{\rm star}\sim 10^{11}$M$_\odot$)
galaxies from the near-infrared Palomar/DEEP-2 survey, we explore which
fraction of these objects has satellites with $0.01<M_{\rm sat}/M_{\rm
central}<1$ (1:100) up to $z=1$ and which fraction has satellites with
$0.1<M_{\rm sat}/M_{\rm central}<1$ (1:10) up to $z=2$ within a projected
radial distance of 100 kpc.  We find that the fraction of massive galaxies
with satellites, after the background correction, remains basically constant and close to
30\% for satellites with a mass ratio down to 1:100 up to $z=1$, and $\sim$15\%
for satellites with a 1:10 mass ratio up to $z=2$. The family of spheroid-like
massive galaxies presents a 2-3 times larger fraction of objects with
satellites than the group of disk-like massive galaxies. A crude estimation of
the number of 1:3 mergers a massive spheroid-like galaxy experiences since
z$\sim$2 is around 2. For a disk-like galaxy this number decreases to $\sim$1.

\end{abstract}


\section{Introduction}

The relevance of major mergers as the main mechanism for the size increase of
the massive (M$_{\rm star}$$\gtrsim$ 10$^{11}$M$_\odot$) galaxies in the last
$\sim$11 Gyr \citep[e.g.,][]{Daddi2005, Trujillo2006, Trujillo2007,
Longhetti2007, Buitrago2008}  has been disfavored observationally
\citep[e.g.,][]{Bundy2009, deRavel2009, Lopez-Sanjuan2010}. This has left room
to a growing consensus that the strong size evolution observed among the
massive galaxies is mainly dominated by the continuous accretion of minor
satellites. However, all the observational evidences compiled so far suggesting
that the minor merging is the main route of galaxy size growth it is only
indirect. The observations that favor the minor merging scenario are: a) a
progressive build-up of the envelopes of the massive galaxies with cosmic time
\citep{Hopkins2009, Bezanson2009, vanDokkum2010, Carrasco2010} and b) a mild
decrease in the velocity dispersion of these galaxies
\citep[e.g.,][]{Cenarro2009, Cappellari2009, Martinez-Manso2011, Newman2010,
vandeSande2011}. Both phenomena agree with a process that do not affect
dramatically their inner regions. Recently, an extra evidence supporting the
merging scenario has been stated: the size evolution of the massive galaxies is
not linked to the age of the stellar population of the galaxies
\citep{Trujillo2011}. All these observations disfavor the puffing-up mechanism
proposed by \citet{Fan2008, Fan2010}, where galaxies grow by the expulsion of
gas by the AGN activity, and give support to the minor merging hypothesis.

On the theoretical side, N-body cosmological simulations as well as
semianalytical models \citep[e.g.,][]{Khochfar-Burkert2006, Naab2009, Oser2011}
show that the expected accretion rate of satellites should be able to produce a
significant increase in the size of the galaxies while at the same time
changing the velocity dispersion only mildly. Estimates of the merger rate
\citep[e.g.,][]{Lopez-Sanjuan2011} using observations are, however, not
straightforward due to the large uncertainties in the determination of the
merging timescales. Nevertheless, a more direct way of confronting simulations
with observations and, consequently, probing the minor merging scenario is to
measure the frequency of satellites found around massive galaxies and quantify
how this fraction changes with cosmic time \citep[e.g.,][]{Newman2011}. Several
papers have calculated this number in the nearby Universe \citep[see
e.g.,][]{Chen2008, Liu2011}. These works show that $\sim$12\% of the massive
galaxies have at least a satellite with a stellar mass
$0.1<M_{sat}/M_{central}<1$ within a projected radius of 100~kpc. These numbers
are in very nice agreement with expectations from $\Lambda$CDM simulations
\citep[see e.g.,][]{Boylan-Kolchin2010}. \citet{deRavel2011} and
\citet{Nierenberg2011} have explored the evolution of the fraction of galaxies
with satellites up to $z \sim 1$ but using mostly samples of central galaxies
less massive than 10$^{11}$M$_\odot$. In this paper we concentrate on the most
massive galaxies and we expand on the previous analysis exploring the fraction
of galaxies with satellites up to $z \sim 2$. To reach our goal we use a large
and  complete sample of massive galaxies up to $z=2$ from \citet{Trujillo2007}.
We probe two different redshift ranges: up to $z=1$, we explore the fraction of
massive galaxies with satellites having $0.01<M_{\rm sat}/M_{\rm central}<1$,
and up to $z=2$, the fraction of massive galaxies with satellites within the
mass range $0.1<M_{\rm sat}/M_{\rm central}<1$.

This letter is structured as follows. In Section~\ref{sec:results} we describe
our sample of massive galaxies and the photometric catalog we have used for
identifying their satellites. Our criteria for selecting satellites as well as
our background estimation methods are explained in Section~\ref{sec:selection}.
Finally, our results are presented in Section~\ref{sec:results}, and a
discussion of our findings in provided in Section \ref{sec:discussion}. In this
paper we adopt a standard $\Lambda$CDM cosmology, with $\Omega_{\rm m} = 0.3$,
$\Omega_{\rm \Lambda} = 0.7$ and H$_0 = 70$ km~s$^{-1}$~Mpc$^{-1}$.

\section{The data}\label{sec:data}

To analyse the evolution with redshift of the fraction of massive galaxies
having satellites, we have used as the reference catalog for the central
galaxies the compilation of massive objects published in \citet{Trujillo2007}
(hereafter T07). This is a homogeneous and large collection of massive galaxies
since $z=2$. Briefly, the sample consists on a total of 831 massive (M$_{\rm
star}$$>$10$^{11}$M$_\odot$) galaxies (of which 35 where identified as AGN and
not used onwards) over 710 arcmin$^2$ in the Extended Groth Strip (EGS). These
objects were K$_s$-band selected in the Palomar Observatory Wide-Field Infrared
(POWIR)/DEEP-2 survey \citep{Bundy2006, Conselice2007}. In total, 372 galaxies
have spectroscopic redshifts \citep{Davis2003}, whereas the remaining redshifts
were obtained photometrically using B, R and I bands from the CFHT
3.6m-telescope, F606W and F814W from the Hubble Space Telescope and J and K$_s$
from the Palomar 5-m telescope. Stellar masses and other derived photometric
parameters were estimated using a Chabrier \citep{Chabrier2003} initial mass
function (IMF). T07 estimated (circularized) half-light radius ($r_{e}$) and
S\'ersic indices n \citep[][]{Sersic1968} for all the galaxies in our sample.

To compile the sample of the satellite galaxies around our massive objects we
have used the EGS IRAC-selected galaxy sample from the Rainbow Cosmological
Database\footnote{\url{https://rainbowx.fis.ucm.es/Rainbow_Database/}}
published by \citet{rainbow} \citep[see also][]{Perez-Gonzalez2008a}. This
database covers an area of $1728~{\rm arcmin}^2$ centred on the EGS and
provides spectral energy distributions (SEDs) ranging from the UV to the MIR
regime plus well-calibrated and reliable photometric redshifts and stellar
masses \citep{rainbow2}. Around 10\% of the galaxies in the Rainbow catalogue
have spectroscopic redshifts. From the Rainbow database we have selected all
the galaxies with $z<2.2$ and an estimated stellar mass $10^8 M_\odot < M <
10^{12} M_\odot$. A total of $\sim$55000 objects were selected in the EGS area
following these criteria. We refer to this resulting sample as the Rainbow
catalog. 

The sample of massive galaxies as well as the Rainbow sample were
cross-correlated using a $1.0\arcsec$ search radius to create a sample of
central galaxies identified in both catalogs. All the massive galaxies in T07
were found in the Rainbow database. The average difference between the
photometric redshifts for the massive galaxies in both samples is $\sim$10\%.
The average stellar mass of our massive sample according to the Rainbow dataset
is 0.9$\times$10$^{11}$M$_\odot$ and 1.7$\times$10$^{11}$M$_\odot$ according to
T07. 

\begin{figure}
\centering
\includegraphics[width=1.10\columnwidth,clip=true]{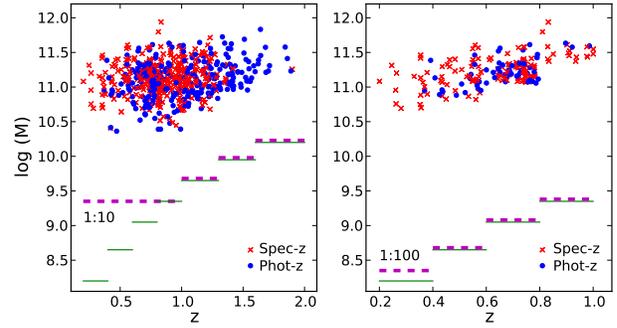}
\caption{Stellar mass vs. redshift for the massive galaxies analised in this
work. Galaxies with spectroscopic redshifts are plotted in red, while galaxies
with photometric estimates are plotted in blue. The left panel shows the
distribution of the massive galaxies used selected for exploring the fraction
of galaxies with 1:10 satellites up to $z=2$. The right panel shows the massive
galaxies used in the study of 1:100 satellites up to $z=1$. The solid green
lines illustrate the stellar mass 75\% completeness limit of the Rainbow
database for the redshift ranges given in \citet{rainbow2}. The magenta dashed
lines the stellar mass cut used in this paper for the different subsamples.}
\label{masssizeplane}
\end{figure}

In order to build a sample of central galaxies with the best estimations of
redshifts and stellar masses we have applied the following rules: a) if a
central galaxy has a spectroscopic redshift determination in Rainbow (348
objects), we have used this redshift plus the stellar mass inferred in that
catalog for these two quantities. Among these galaxies, there were 8 objects
with spectroscopic redshifts in both samples with high discrepancies in the
stellar mass estimations from both catalogs. We reject from our sample such
dubious cases. b) If no spectroscopic redshift is found on the Rainbow database
but on the T07's sample (37 galaxies) we use the values of redshift and stellar
masses from that catalog. c) Finally, if no spectroscopic redshift is found in
any of the two catalogs we have used only those objects where the photometric
redshift determination is robust (317 objects). This means that we have
compared the two independent photo-z estimations found in T07 and
\citet{rainbow} and we have only taken those objects where the photometric
redshifts disagree less than $\Delta z_{\rm phot} = 0.070$ for $0.0<z<0.5$,
$\Delta z_{\rm phot} = 0.061$ for $0.5<z<1.0$, and $\Delta z_{\rm phot} =
0.083$ for $1.0<z<2.5$ \citep[typical quality of the photo-z's in the Rainbow
catalog in EGS obtained by comparing them with spec-z's,][]{rainbow2}. This
removes 94 galaxies. For consistency with the sample of satellite galaxies, for
these 317 objects we take the stellar masses and photometric redshift from the
Rainbow catalog. After this selection, the number of objects in the final
sample of massive galaxies is 694, of which 317 have photometric redshifts from
Rainbow and 377 have spectroscopic redshifts (340 from the Rainbow catalog and
37 from T07).

\begin{figure*}
\begin{minipage}{0.33\textwidth}
\includegraphics[width=0.90\columnwidth,clip=true]{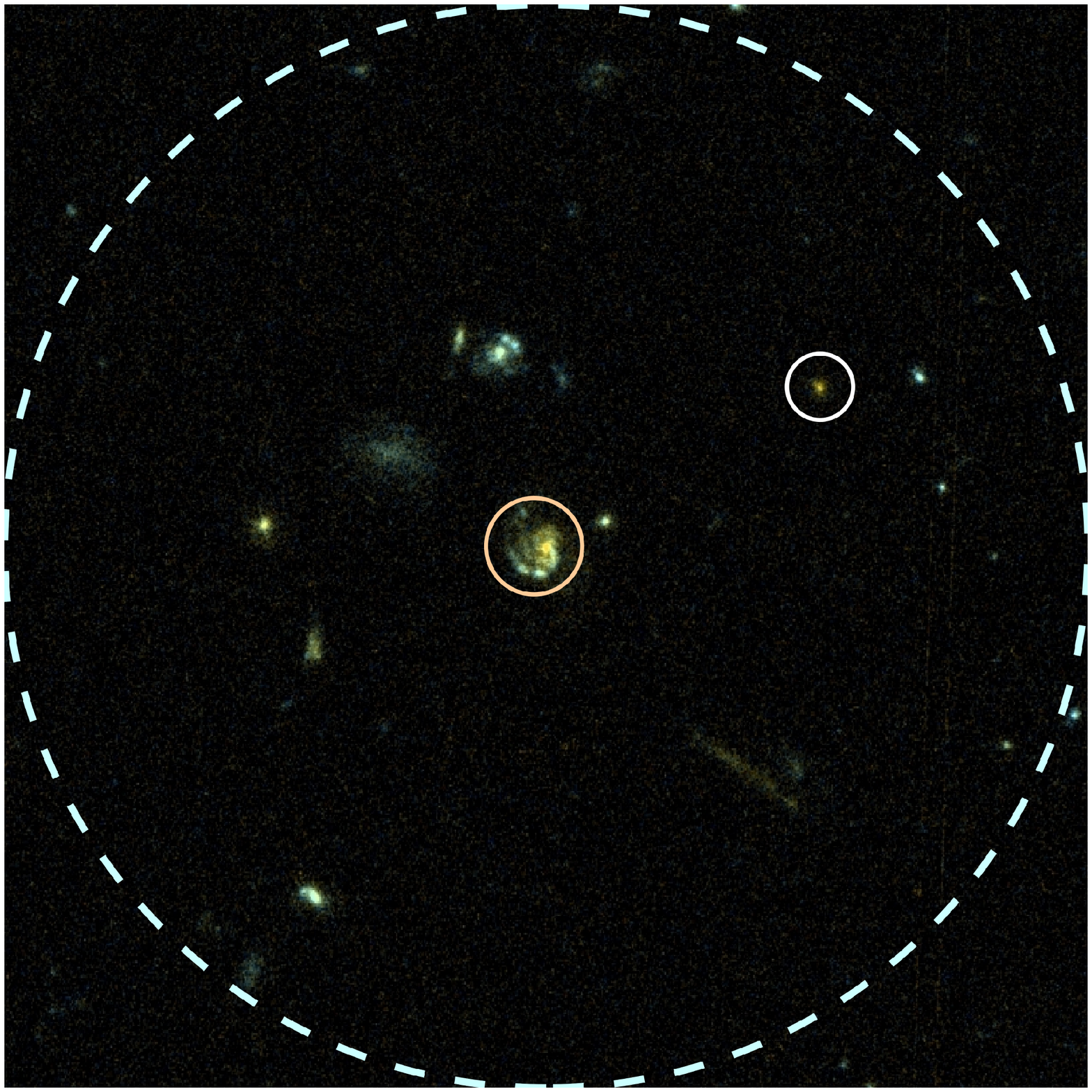}
\end{minipage}
\hfill
\begin{minipage}{0.33\textwidth}
\includegraphics[width=1.00\columnwidth,angle=-90,clip=true]{irac123359.SED_fit_specz.ps}
\end{minipage}
\hfill
\begin{minipage}{0.33\textwidth}
\includegraphics[width=1.00\columnwidth,angle=-90,clip=true]{irac123191_1.SED_fit_specz.ps}
\end{minipage}
\caption{{\it: Left panel:} ACS color image of the massive galaxy IRAC123359
(in the centre) at $z=1.17$ with a satellite galaxy (IRAC123191-1) that meets
the selection criteria used in this work enclosed by a white circle. A circle
of radius 100 kpc is plotted with a dashed line. {\it Central and right
panels:} Spectral energy distributions for both the massive (central panel) and
the satellite (right) galaxies. These panels also include the redshift and
stellar mass estimates according to the Rainbow database.} 
\label{example_satellites} 
\end{figure*}

A final cut in the number of galaxies of our main sample is required to assure
that the fraction of galaxies with satellites along our explored redshift range
is not biased by the stellar mass completeness limit of the Rainbow database.
The stellar mass limit (75\% complete) of the Rainbow database at each redshift
is provided in \citet[][see their Fig.~4]{Perez-Gonzalez2008a}. In the redshift
range $0<z<2$ we have selected only those massive galaxies whose stellar masses
are 10 times larger than the completeness limit at each redshift. There are 629
galaxies (with a mean stellar mass of $M =1.3\times10^{11} M_\odot$ for this
sample) that meet these criteria. On doing that we are secure that we can
explore within the Rainbow catalog satellites down to 1:10 mass ratio of the
central galaxy in the range $0<z<2$. For the same reason, this exercise is done
up to $z=1$ but this time only selecting those central galaxies with a stellar
mass 100 times above the mass limit. This cut leaves us in the redshift range
$0<z<1$ with 194 massive galaxies (with a mean stellar mass of $M =
1.7\times10^{11} M_\odot$. The stellar masses and the redshifts for the
central galaxies studied in this work are illustrated in
Fig.~\ref{masssizeplane}.

\section{Selection criteria}\label{sec:selection}

To identify the  satellite galaxies around our central objects we have applied
the following procedure: (1) we identify all the galaxies in the Rainbow
catalog which are within a projected radial distance to our central galaxies of
$R_{\rm search}$=100 kpc (corresponding to 0.3 and 0.2 arcmin for $z=0.5$
and $z=2$, respectively); (2) the difference between their photometric redshifts
and the redshift of the central galaxies is lower than the 1$\sigma$
uncertainty in the estimation of the photometric redshifts of the Rainbow
database (i.e., $\Delta z_{\rm phot} = 0.070$ for $0<z<0.5$, $\Delta z_{\rm
phot} = 0.061$ for $0.5<z<1$, and $\Delta z_{\rm phot} = 0.083$ for
$1<z<2.5$); and (3) the stellar mass of these objects should be within
$0.1<M_{\rm sat}/M_{\rm central}<1.0$ for the galaxies in the range $0<z<2$,
and within $0.01<M_{\rm sat}/M_{\rm central}<1.0$ for the galaxies in the range
$0<z<1$. An example of satellite galaxy satisfying the above criteria is shown
in Fig.~\ref{example_satellites}. Finally, we consider different redshift bins
(see Table~\ref{table_results}) to explore the evolution of the fraction,
$F_{\rm sat}$, of massive galaxies with satellites. The width of these bins
were chosen to include a similar number of massive galaxies in each bin and
have a similar statistics among them.

\begin{figure}
\centering
\includegraphics[angle=-90, width=0.90\columnwidth]{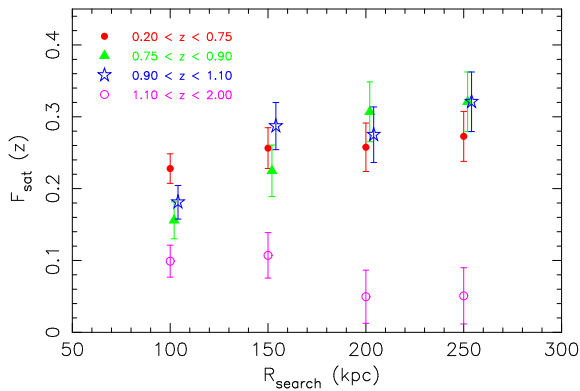}
\caption{Fraction of massive galaxies having satellites within different
projected radial distances (search radius, $R_{\rm search}$) in the mass range
$0.1<M_{\rm sat}/M_{\rm central}<1.0$ for the different redshift bins studied in this
work.}

\label{search_radius}
\end{figure}

We adopted a search radius of 100~kpc. This radius is a compromise between
having a large area for finding a significant number of satellite candidates
that are gravitationally bound to our central massive galaxies but not as large
as to be severely contaminated by background objects. In any case, we have
explored what is the effect on our measurements if we select larger radii of
exploration. We computed the fraction of massive galaxies with satellites for
different search radii ($R_{\rm search}$=100, 150, 200 and 250 kpc). The
results of this experiment in the mass range $0.1<M_{\rm sat}/M_{\rm
central}<1$ are shown in Fig.\ref{search_radius}. The numbers presented here
are corrected of background contamination as it will be explained later. As it
is expected, we detect an increasing number of massive galaxies with satellites
as we expand the search radius $R_{\rm search}$. The only exception is the
redshift range $1.1 < z < 2.0$ where the fraction of massive galaxies with
satellites is constant within the error bars. It is worth noting that, in
general, beyond $R_{\rm search}$=150~kpc there is not a net increase in the
fraction of massive galaxies with satellites. Moreover, our results are
basically unchanged if we use a search radius of 100~kpc or 150~kpc. For this
reason, on what follows we will present the results based on a search radius of
100~kpc as our simulations show that this case is affected by the background
contamination a factor of $\sim2$ less than the 150~kpc case.

Similarly to the selection of the search radius, we have restricted our
potential satellite galaxies to have a redshift difference with the central
galaxy not larger than the 1$\sigma$ uncertainty in the estimation of the
photometric redshifts of the Rainbow database. Larger redshift differences
could be used to include more potential candidates but this is transformed also
into a larger background contribution to our measurements. For instance, we
estimated how the fraction of massive galaxies with satellites changed when
using 2$\sigma$ uncertainty in the estimation of the photometric redshifts
instead of 1$\sigma$. As expected, we found a slight increase ($\lesssim$30\%)
on the fraction but also our error bars increased (by a 50\%) by the larger
amount of background contamination. As these change do not alter our main
results but increase our error bars, we have used the 1$\sigma$ criteria
explained in this paper.

\subsection{Background estimation}\label{sub:background}

Despite we have used photometric redshift information to select our potential
satellite galaxies, there is still a fraction of objects that satisfy the above
criteria but are not gravitationally bound to our massive galaxies. These
objects are counted as satellites because the uncertainties on their redshift
estimates include them within our searching redshift range. These foreground
and background objects (hereafter we will use the term background to refer to
both of them) constitute the main source of uncertainty in this kind of
studies. Consequently, it is key to estimate accurately the background
contamination in order to statistically subtract its contribution from the
fraction of galaxies with observed satellites.

To estimate the fraction of background sources that contaminates our satellite
samples  we have run a set of simulations. This method consists on placing a
number of mock massive galaxies (equal to the number of our central galaxies)
randomly through the volume of the catalog. To match our observed redshift
distribution we assure that in our simulations, the number of mock galaxies
that are within each redshift bin is the same than in our observed sample. Once
we have placed our mock galaxies through the catalog, we count which fraction
of these mock galaxies have satellites around them taking into account our
criteria of redshift and distances explained above. This procedure is repeated
two million times to have a robust estimation of the fraction of mock galaxies
with satellites. We call this average fraction $S_{\rm simul}$.  Additionally,
these simulations allowed us to estimate the scatter in the fraction of
galaxies that have contaminants. We use this scatter as an estimation of the
error of our real measurements. We consider then this fraction to be
representative of the background affecting our real central sample. The
galaxies in our mock samples keep fixed the parameters of the massive galaxies
(i.e., stellar masses and S\'ersic indices).

Taking into account that the observed fraction of galaxies with satellites,
$F_{\rm obs}$, is the sum of the fraction of galaxies with real satellites,
$F_{\rm sat}$, plus the fraction of galaxies which have not satellites but
are affected by contaminants ($1-F_{\rm sat}$)$\times$$S_{\rm simul}$ we
arrive to the following expression:
\begin{equation}\label{formula_frac}
F_{\rm sat} = \frac{F_{\rm obs} - S_{\rm simul}}{1-S_{\rm simul}}.
\end{equation}

The results of our simulations are shown in Table~\ref{table_results}. From
these simulations we see that the fraction $S_{\rm simul}$ of massive galaxies
we expect to be contaminated by false satellites (using our searching criteria)
is $\sim$10\% for $0.1<M_{\rm sat}/M_{\rm central}<1.0$ and $\sim$25\% for
$0.01<M_{\rm sat}/M_{\rm central}<1.0$.

\begin{figure*}
\centering
\includegraphics[angle=-90, width=0.95\textwidth]{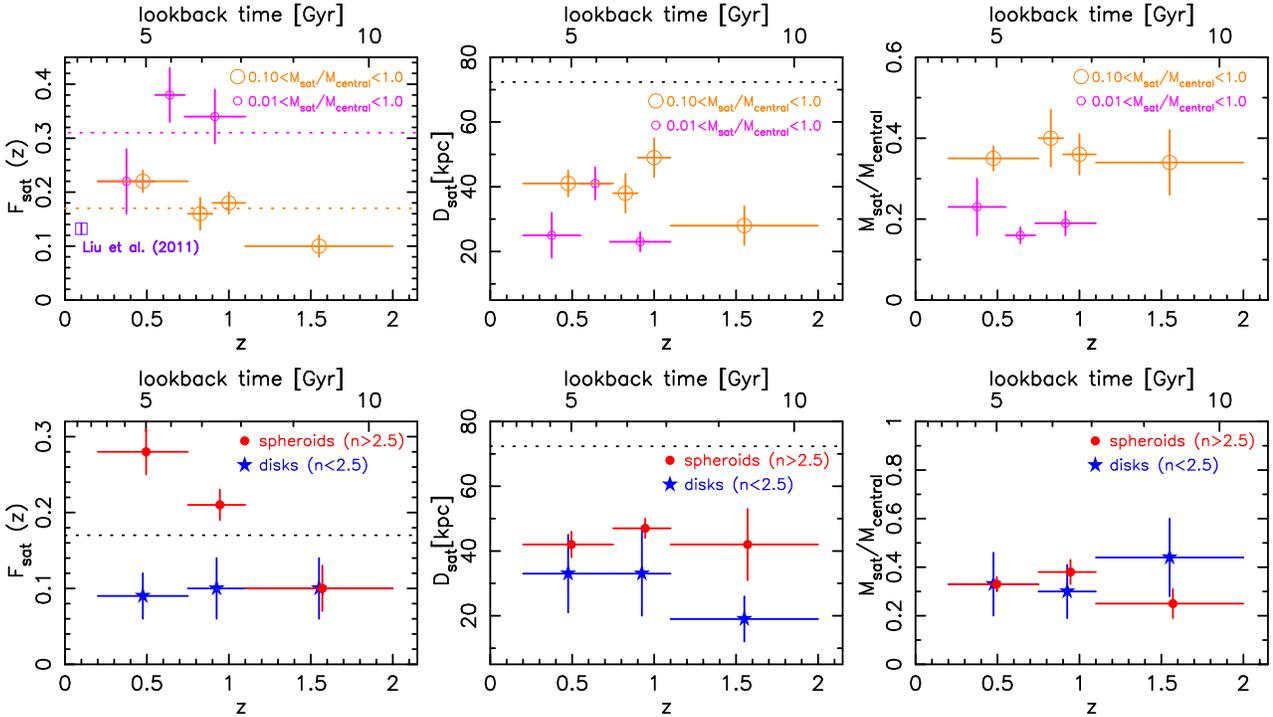}
\caption{Fraction of massive galaxies having satellites (and their properties)
within a projected radial distance of 100 kpc for different redshift bins.
Upper panels show the fraction of massive galaxies with satellites in the mass
range $0.1<M_{\rm sat}/M_{\rm central}<1.0$ (orange dots) and the fraction of
massive galaxies with satellites in the mass range $0.01<M_{\rm sat}/M_{\rm
central}<1.0$ (magenta dots). In the bottom panels, we explore the fraction of
massive galaxies with satellites in the mass range $0.1<M_{\rm sat}/M_{\rm
central}<1.0$ when our sample is split depending on the S\'ersic index
(morphology) of the galaxies (red dots for $n>2.5$ -spheroids- and blue stars
for $n\le2.5$ -disks). The horizontal dotted lines in the upper left panel
correspond to the average fitted values to our findings. {\it Central panels:}
Mean projected distances where the satellites are found after statistical
correction of the background. The dashed line indicates the average
projected distance obtained for mock satellites in the simulations (i.e. this is the expected distance if the
satellites where an artifact of the background contamination).
{\it Right panels:} Mean mass ratios between the satellites and their massive
galaxies. The horizontal bars indicate the range of redshifts considered for
each measurement. For clarity, we have slightly shifted the data corresponding
to the spheroids-like objects.}

\label{figures}
\end{figure*}

\subsubsection{Clustering effects}\label{sec:clustering}

It is well known that massive galaxies, particularly in the nearby Universe,
tend to populate regions which are overdense compared to the average density of
the Universe. This implies that there is an excess of probability (which we
will term as clustering) of finding galaxies that could be misidentified as
satellites of our main targets. It is worth noting that this probability excess
is not related to the accuracy of our redshift estimations. Even with all the
redshifts measured spectroscopically, the effect of clustering will be equally
relevant in our estimates as this effect is inherent to our inability of
measuring real distances but distances inferred by recessional velocities. In
massive cluster of galaxies, with velocity dispersion of
$\sim1000$~km~s$^{-1}$, this will limit our accuracy on estimating real galaxy
associations. 

Being the clustering a local effect, ideally one would like to measure its
influence as closer as possible to the central galaxy. In practice, this is
done by measuring the amount of satellite candidates in different annuli beyond
our search radius \citep{Chen2006, Liu2011}. We will call to the fraction of
massive galaxies having satellites in these annuli as $S_{\rm cluster}$. This
fraction measures both the effect of the background contamination plus the
excess over this background due to the clustering. This method has the
disadvantage, compared to the simulations that we have conducted above, that is
statistically more uncertain. $S_{\rm cluster}$ can be measured only around our
massive galaxies and this number is relatively small. For this reason, $S_{\rm
cluster}$ is determined with an error larger than $S_{\rm simul}$. We count the
satellites in 9 different annuli  in the radial range $100<R<330$~kpc (the size
of each annuli was selected to contain the same area than the searching area
within 100~kpc). We find, as expected, that the number of satellites decreases
in the outer annuli, reaching asymptotically (within the errors) the values we
get using the first background estimation method. However, in general, and
particularly for the lower redshift bins, the number of detected satellites is
higher for the inner annuli than in the random case, and therefore, the
clustering is not negligible. As we noted before, the detection of satellites
does not increase at $R_{\rm search}>150$~kpc. For this reason, and as a
compromise between proximity to the massive galaxies and having enough
statistics, we have used the average detections of satellites in the two annuli
closer to R=150~kpc ($173 < R < 200$~kpc and $200 < R < 224$~kpc) to estimate
the effect of the clustering. The uncertainty at measuring $S_{\rm
cluster}$ is not straightforward to calculate and we have decided to estimate
that value summing quadratically the background uncertainty measured in the
simulations estimating $S_{\rm simul}$ plus the dispersion between the two
different radial annuli used in the clustering determination. 

The significance of the clustering is quantified in Table~\ref{table_results}.
We find that above $z>1$ the clustering is playing a minor role as $S_{\rm
cluster}$ and $S_{\rm simul}$ are very much alike within the errors. However,
at $z<1$, $S_{\rm cluster}\sim1.5\times S_{\rm simul}$. As it is expected, the
effect of the clustering is more relevant at lower redshifts. At high
redshifts, the overdensities are less significant as the large scale structures
would be still not completely formed.

\section{Results}\label{sec:results}

In Table~\ref{table_results} we summarize the results obtained in this work.
For each redshift bin, we present the fraction of galaxies with satellites
initially found for our sample of massive galaxies, $F_{\rm obs}$, the
background estimate $S_{\rm simul}$ derived from the mock catalogues, and the
final fraction of massive galaxies with satellites, $F_{\rm sat}$, after the
correction of the background contamination with Eq.~\ref{formula_frac}. The
associated errors correspond to the standard deviation from the measurements
obtained in the mock catalogues as explained in the previous section. In
addition, we include the expected contamination due to the clustering estimate,
$S_{\rm cluster}$, and the fraction of massive galaxies with satellites after
this correction $F_{\rm cluster}$.

Our results are also illustrated in Fig.~\ref{figures}. Our main result is seen
in the upper left panel of Fig.~\ref{figures}: the fraction of massive galaxies
with satellites, within a projected radial distance of 100~kpc, in the range
$0.1<M_{\rm sat}/M_{\rm central}<1$ remains basically constant ($17\pm3$\%) in
the redshift interval $0<z<2$.  To have a $z=0$ comparison, we have added the
measurement from \citet{Liu2011} using the SDSS sample. They find that at $z=0$
the fraction of massive galaxies with satellites in the mass range and
projected radius explored here is very similar. In the same panel, we show the
same analysis up to $z=1$ for satellite galaxies with $0.01<M_{\rm sat}/M_{\rm
central}<1.0$. Although a little bit noisier due to the lower statistics, our
findings agree with a relative constant fraction ($31\pm6$\%) of massive
galaxies having such type of satellites.

Our sample of massive galaxies is large enough that we can explore whether the
fraction of massive galaxies with satellites in the range $0.1<M_{\rm
sat}/M_{\rm central}<1$ depends on the morphology of the massive galaxy. We
have used the S\'ersic index as a proxy to the galaxy morphology. In the nearby
universe, galaxies with $n < 2.5$ are mostly disc-like objects, whereas
galaxies with $n > 2.5$ are mainly spheroids \citep[e.g.,][]{Andredakis1995,
Blanton2003c, Ravindranath2004}. We have used the published S\'ersic indexes
provided by T07 to separate our galaxies. We illustrate our results in the
bottom left panel of Fig.~\ref{figures}. There is a hint that massive galaxies
with spheroid-like morphologies tend to have a larger fraction (a factor of
2-3) of galaxies with satellites than disk-like massive objects. This result is
more prominent at low redshift where the clustering of the massive spheroid
population could be an issue. 

\begin{figure} 
\centering 
\includegraphics[angle=-90,width=0.90\columnwidth]{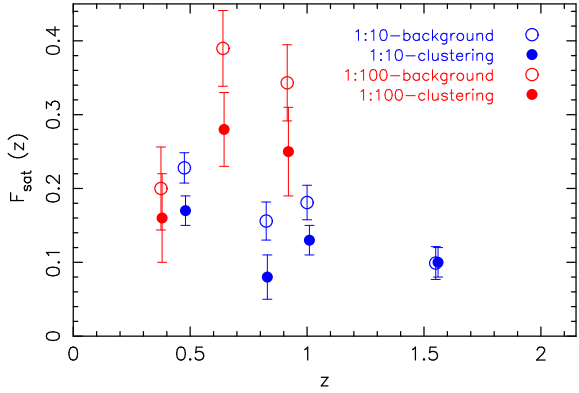} 
\vspace{0.5cm}

\includegraphics[angle=-90,width=0.90\columnwidth]{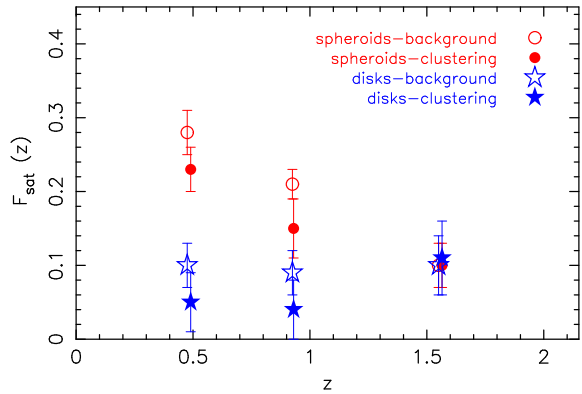} 

\caption{Fraction of massive galaxies with satellites after correction  with
the background contribution computed from the simulations (open symbols)  and
when the clustering estimation is used (filled symbols). {\it Top panel:} The
fraction of massive galaxies with satellites is separated in two groups:
$0.1<M_{\rm sat}/M_{\rm central}<1.0$ (blue triangles) and $0.01<M_{\rm
sat}/M_{\rm central}<1.0$ (red squares). {\it Bottom panel:} Fraction of
massive galaxies in the mass range $0.1<M_{\rm sat}/M_{\rm central}<1.0$
depending on the S\'ersic index (morphology) of the central galaxies.}

\label{clustering_effects}
\end{figure}

We can repeat the same exercise but this time using the clustering correction
(which contains also the background effect) to explore how  our results depend
on this effect. The comparison between the two types of corrections are shown
in Fig.~\ref{clustering_effects}. In general, the correction due to the
clustering decreases the fraction of massive galaxies which contains
satellites. This is now $12\pm2$\% in the redshift interval $0<z<2$ for
$0.1<M_{\rm sat}/M_{\rm central}<1$ and $23\pm4$\% for $0.01<M_{\rm sat}/M_{\rm
central}<1.0$ up to $z=1$.

\begin{table*}
\centering

\caption{Fraction of massive galaxies with satellites at different redshifts.
For each redshift range we present the number of massive galaxies $N_{\rm
central}$ in each bin (number of galaxies with spectroscopic redshifts in
brackets), the observed fraction of massive galaxies with satellites  $F_{\rm
obs}$, the estimate of the background contamination $S_{\rm simul}$, and the
estimate of the clustering effect $S_{\rm cluster}$.  Finally, we present the
final fraction of massive galaxies with satellites when i) the correction for
the background contamination ($F_{\rm sat}$)  or ii) the clustering effect
($F_{\rm cluster}$) is applied.}

\begin{tabular}{lcccccc}
\hline\hline
Redshift range    & $N_{\rm central}$  & $F_{\rm obs}$ & $S_{\rm simul}$ & $S_{\rm cluster}$ & $F_{\rm sat}$  & $F_{\rm cluster}$ \\
                  & ($N$ with spec z) &               &               &      &      &     \\
\hline
\multicolumn{7}{l}{\bf All galaxies}       \\
\hline          
\multicolumn{7}{l}{$0.10 < M_{\rm sat}/M_{\rm central} < 1.00$ }\\
$0.20 < z < 0.75$ & 197 (130) & $0.29$ & $0.09 \pm 0.02$ & $0.15 \pm 0.02$ & $0.22 \pm 0.02$ & $0.17 \pm 0.02$  \\
$0.75 < z < 0.90$ & 129 (76)  & $0.24$ & $0.10 \pm 0.03$ & $0.17 \pm 0.03$ & $0.16 \pm 0.03$ & $0.08 \pm 0.03$  \\
$0.90 < z < 1.10$ & 142 (99)  & $0.25$ & $0.08 \pm 0.02$ & $0.13 \pm 0.03$ & $0.18 \pm 0.02$ & $0.13 \pm 0.02$  \\
$1.10 < z < 2.00$ & 161 (55)  & $0.18$ & $0.09 \pm 0.02$ & $0.09 \pm 0.03$ & $0.10 \pm 0.02$ & $0.10 \pm 0.02$  \\
 & & & & & \\ 
\hline          
\multicolumn{7}{l}{$0.01 < M_{\rm sat}/M_{\rm central} < 1.00$ }\\
$0.20 < z < 0.55$ &  51 (40) & $0.37$ & $0.20 \pm 0.06$ & $0.24 \pm 0.06$ & $0.22 \pm 0.06$ & $0.16 \pm 0.06$  \\
$0.55 < z < 0.73$ &  70 (42) & $0.53$ & $0.24 \pm 0.05$ & $0.36 \pm 0.05$ & $0.38 \pm 0.05$ & $0.28 \pm 0.05$  \\
$0.73 < z < 1.10$ &  73 (53) & $0.52$ & $0.27 \pm 0.05$ & $0.36 \pm 0.06$ & $0.34 \pm 0.05$ & $0.25 \pm 0.06$  \\
\hline                                                                                 
\multicolumn{7}{l}{${\bf Spheroid-like}$ ${\bf (n > 2.5)}$ ${\bf galaxies}$}\\
\hline                                                                                 
\multicolumn{7}{l}{$0.10 < M_{\rm sat}/M_{\rm central} < 1.00$ }  \\
$0.20 < z < 0.75$ & 137 & $0.34$ & $0.09 \pm 0.03$ & $0.16 \pm 0.03$ & $0.28 \pm 0.03$ & $0.23 \pm 0.03$ \\
$0.75 < z < 1.10$ & 176 & $0.27$ & $0.08 \pm 0.02$ & $0.14 \pm 0.04$ & $0.21 \pm 0.02$ & $0.15 \pm 0.04$ \\
$1.10 < z < 2.00$ &  85 & $0.18$ & $0.08 \pm 0.03$ & $0.08 \pm 0.03$ & $0.10 \pm 0.03$ & $0.10 \pm 0.03$ \\
\hline                                                                                 
\multicolumn{7}{l}{${\bf Disk-like}$ ${\bf (n < 2.5)}$ ${\bf galaxies}$}  \\
\hline                                                                                 
\multicolumn{7}{l}{$0.10 < M_{\rm sat}/M_{\rm central} < 1.00$ }  \\
$0.20 < z < 0.75$ &  60 & $0.18$ & $0.10 \pm 0.03$ & $0.15 \pm 0.04$ & $0.09 \pm 0.03$  & $0.05 \pm 0.04$ \\
$0.75 < z < 1.10$ &  95 & $0.19$ & $0.09 \pm 0.03$ & $0.16 \pm 0.04$ & $0.10 \pm 0.04$  & $0.04 \pm 0.04$ \\
$1.10 < z < 2.00$ &  76 & $0.18$ & $0.09 \pm 0.04$ & $0.09 \pm 0.05$ & $0.10 \pm 0.04$  & $0.11 \pm 0.05$ \\
\hline
\end{tabular}
\label{table_results}
\end{table*}

\subsection{Robustness of the results}

The results presented in this paper are the product of combining two different
datasets: the T07 sample of massive galaxies and the Rainbow catalogue. In
addition, we have used photometric redshifts and, when available (54\% of the
time), spectroscopic redshifts. We have checked how robust are our results to
the use of more homogeneous dataset, as using only a sample of massive galaxies
with spectroscopic redshifts or, to base our full analysis only on the Rainbow
database. 

\begin{figure} \centering
\includegraphics[angle=-90,width=0.90\columnwidth]{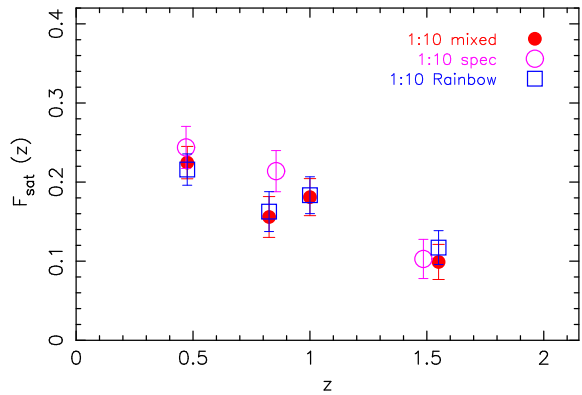}
\caption{Testing the robustness of the fraction of massive galaxies with
satellites in the mass range $0.1<M_{\rm sat}/M_{\rm central}<1.0$ for
different datasets: our original sample described in Sect.~2 (red points), a
purely spectroscopic sample (magenta circles) and a sample based only on the
Rainbow catalogue (blue squares).}
\label{comparison_samples} 
\end{figure}

In our first test, we used only the central galaxies in our sample with
spectroscopic redshifts and counted which fraction has satellites following the
same procedure explained above and after correcting for the background.  The
width of the redshift bins for this sample were again chosen to include a
similar number of massive galaxies and have a similar statistics to compare
with the other samples. The output of this test is shown in
Fig.\ref{comparison_samples}. We get, for the case of $0.1<M_{\rm sat}/M_{\rm
central}<1$ up to $z=2$, an average fraction of $19\pm4$\%. It can be seen that
this result is in full agreement with our previous estimation for this
quantity. In a second test, we have taken the redshifts and stellar masses only
from the Rainbow catalog to check whether there are systematic effects due to
the use of combining different samples. We get, in this case, a fraction of
$17\pm2$\%. Again, this result agrees perfectly with the original estimate. We
conclude, accordingly, that our results are robust to both the use of
spectroscopic redshifts only and to the mixing of different datasets.

Another test that we have conducted is to check whether our results are robust
to a change in the stellar mass limit at selecting our massive galaxies. As it
is illustrated in Fig. \ref{masssizeplane}, the galaxies at higher redshifts
are slightly more massive than the bulk of objects at lower redshift to
guarantee that we can study satellites above a given mass ratio along the full
redshift range. We have checked how our results change if we select only
massive galaxies with stellar masses above 2$\times$10$^{11}$ M$_\odot$. We
have done this exercise for the case $0.1<M_{\rm sat}/M_{\rm central}<1$ up to
$z=2$. With the new mass limit, we get an average fraction of $\sim23$\%. This
is in good agreement with our original estimation for this fraction. Again,
increasing the stellar mass limit does not alter substantially the fraction of
massive galaxies with redshift.

\subsection{Properties of the satellite galaxies}

In addition to counting which fraction of the massive galaxies have satellites,
we can also estimate the average projected radial distances of these satellites
and the average mass ratios between the satellites and the massive objects. To
estimate these quantities properly, we need to correct statistically by the
effect of the contaminants. This can be done using the following expression:
\begin{equation}\label{correction_parameters} < Q_{\rm sat} > = \frac{F_{\rm obs}}{F_{\rm sat}} <Q_{\rm obs}> - \frac{S_{\rm
simul}}{F_{\rm sat}}  <Q_{\rm simul}> \end{equation} where $<Q_{\rm obs}>$ is the observed mean value of the property $Q$ (i.e. the
projected radial distance or the mass ratio), $<Q_{\rm simul}>$ is the mean
value obtained from the mock massive galaxies (i.e. the values that are found
for the contaminants) and $<Q_{\rm sat}>$ is the value after the correction.

The mean projected radial projected distances (in kpc) of the satellites and
their mean mass ratios $M_{\rm sat}/M_{\rm central}$ are compiled in
Table~\ref{table_properties} and shown in Fig.~\ref{figures}. We plot with a
dashed line the average projected distance ($\sim72$~kpc) of the background
galaxies detected as fake satellites in the simulations. After correcting by
the effect of the contaminants, we find that our satellite galaxies are at a
typical projected radial distance of $\sim$40~kpc. This value is well below the
expectation from a random distribution, suggesting that the satellites are
gravitationally bounded to their central galaxies. This average distance seems
to be pretty much independent (within the errors) of the satellite mass, the
morphological type of the central galaxy and the redshift of the system.

Finally, we show in the right panels of Fig.~\ref{figures} the mean mass ratio
$M_{\rm sat}/M_{\rm central}$ in each redshift bin after the statistical
correction. We find that $M_{\rm sat}$ is around 0.36 $M_{\rm central}$ when we
explore satellite galaxies within a mass ratio of $0.1 < M_{\rm sat}/M_{\rm
central} <1$ (this value is 0.28 when we use the clustering correction). If we
explore down to a mass ratio of 0.01 then $M_{\rm sat}$ is around 0.15 $M_{\rm
central}$ (0.14 correcting by the clustering effect). It is worth noting that
in both cases, the mean masses of our satellites are over $10^{10}M_\odot$ and
consequently we are detecting satellites with large masses. When we split the
sample depending on their S\'ersic indices (bottom right panel in
Fig.~\ref{figures}), there are not significant differences, within the errors,
between both samples. Again we find that the satellites of massive galaxies are
similar in their mean properties (distances and mass ratios) independently of
the morphological type of their central galaxies and redshift.

\begin{table}
\centering
\caption{Mean projected radial distances between the satellites and the central galaxies
and their mean mass ratios ($M_{\rm sat}/M_{\rm central}$). These values
correspond to the case where only the background correction has been applied.}
\begin{tabular}{lcc}
\hline\hline
Redshift range     &  Radial Distance &  $M_{\rm sat}/M_{\rm central}$ \\
                  &  (kpc)    &                             \\
\hline
{\bf All galaxies}&     &              \\
\hline          
\multicolumn{3}{l}{$0.10 < M_{\rm sat}/M_{\rm central} < 1.00$ } \\
$0.20 < z < 0.75$  & $41 \pm 4$ & $0.35 \pm 0.03$ \\
$0.75 < z < 0.90$  & $38 \pm 6$ & $0.40 \pm 0.07$ \\
$0.90 < z < 1.10$  & $49 \pm 6$ & $0.36 \pm 0.05$ \\
$1.10 < z < 2.00$  & $28 \pm 6$ & $0.34 \pm 0.08$ \\
 & & \\ 
\hline          
\multicolumn{3}{l}{$0.01 < M_{\rm sat}/M_{\rm central} < 1.00$ } \\
$0.20 < z < 0.55$ & $25 \pm 7$ & $0.23 \pm 0.07$ \\
$0.55 < z < 0.73$ & $41 \pm 5$ & $0.16 \pm 0.02$ \\
$0.73 < z < 1.00$ & $23 \pm 3$ & $0.19 \pm 0.03$ \\
\hline                                                                                 
\multicolumn{3}{l}{${\bf Spheroid-like}$ ${\bf (n > 2.5)}$ ${\bf galaxies}$}    \\
\hline                                                                                 
\multicolumn{3}{l}{$0.10 < M_{\rm sat}/M_{\rm central} < 1.00$ } \\
$0.20 < z < 0.75$  & $42 \pm 4$ & $0.33 \pm 0.03$ \\
$0.75 < z < 1.10$  & $47 \pm 3$ & $0.38 \pm 0.05$ \\
$1.10 < z < 2.00$  & $42 \pm 11$ & $0.25 \pm 0.06$ \\
\hline                                                                                 
\multicolumn{3}{l}{${\bf Disk-like}$ ${\bf (n < 2.5)}$ ${\bf galaxies}$}  \\
\hline                                                                                 
\multicolumn{3}{l}{$0.10 < M_{\rm sat}/M_{\rm central} < 1.00$ } \\
$0.20 < z < 0.75$ & $33 \pm 12$ & $0.33 \pm 0.13$ \\
$0.75 < z < 1.10$ & $33 \pm 13$ & $0.30 \pm 0.11$  \\
$1.10 < z < 2.00$ & $19 \pm 7$ & $0.44 \pm 0.16$  \\
\hline
\end{tabular}
\label{table_properties}
\end{table}


\section{Discussion and conclusions}\label{sec:discussion}

The results of this paper support a picture where the fraction of massive
($M_{\rm star}\sim 10^{11}M_\odot$) galaxies with satellites, within
a projected radius of 100~kpc, has not changed with time since $z\sim2$. This
fraction remains around $\sim15$~\% for galaxies with satellites with mass
$M_{\rm star}\gtrsim 10^{10}M_\odot$ and around $\sim30$\% if we explore
satellites with masses $M_{\rm star}\gtrsim 10^{9}M_\odot$ up to $z=1$. 

Interestingly, we find a hint that the fraction of massive galaxies with
satellites is larger (a factor of around 2 to 3) for those galaxies with
spheroid-like morphologies than for galaxies with disk-like appearance (much
evident for $z\lesssim1.1$). This fact could be linked to the different size
growth we observe for these two types of objects as cosmic time increases. In
fact, spheroid galaxies are known for growing more dramatically in size since
$z\sim3$ than disk galaxies \citep[see e.g., T07][]{Buitrago2008}. It could be
also possible that the fraction of satellites difference between spheroid and
disk-like galaxies would be just an effect of the clustering, more relevant at
lower redshifts (Sect.~\ref{sec:clustering}). However, this difference remains
even when this effect is taken into account (Fig.~\ref{clustering_effects}). We
remark, however, that it is difficult to correct the clustering effect
accurately.  With the present dataset, a mild redshift evolution of the
fraction of spheroid-like galaxies with satellites can not be excluded.

Due to the enormous uncertainty on the merging timescales
\citep[e.g.,][]{Lotz2011}, it is beyond the scope of this work to estimate a
robust merger rate associated to our measurements. Nonetheless, we can make a
crude estimation on the number of mergers a massive galaxy experiences since a
given z according to the following expression: $N_m = T(z)\times F_{\rm
sat}/\tau_{\rm m}$, where $T(z)$ is the interval of cosmic time since a given
$z$ to now, and $\tau_{\rm m}$ is the merging timescale of the satellite within
a given radius. For each massive galaxy at $z=2$ and assuming $\tau_{\rm
m}\sim$1.5~Gyr \citep[e.g.][]{Lotz2011}, we would expect that the number of
mergers with mass ratio around 1:3 would be $\sim$1 ($\sim$2 in the case of an
spheroid-like galaxy) since that epoch. For a massive galaxy at $z=1$, we would
expect that the number of mergers with mass ratio around 1:6 would be $\sim$1.5
since that redshift. Again, these numbers are uncertain and very much dependent
on the exact merging timescale which is a function of the baryonic mass ratio
and the model used to estimate this quantity
\citep[e.g.,][]{Bluck2009,Conselice2009,Lotz2011, Man2011}. These numbers of
mergers, however, are slightly lower (although the exact amount is difficult to
quantify) than the expected number of mergers obtained using theoretical
recipes for the size increase of a galaxy after a merger
\citep[see][]{Trujillo2011}.  Recently, \citet{Bluck2011} find that a massive
galaxy ($M_{\rm star}>10^{11}M_\odot$) will experience on average  $N_m = (1.1
\pm 0.2)/\tau_{\rm m}$ minor mergers over the redshift range $z=1.7-3$.  This
would mean a final $N_m \sim 1$ using the $\tau_{\rm m}$ considered in our
work. If this is confirmed, it will point out to the possibility that the
merging activity at those redshifts would be higher than at lower redshifts.
When extrapolated in redshift, they find a total final number of minor mergers
of $N_m = (4.5 \pm 2.9)/\tau_{\rm m}$ from $z=3$, although once more, the large
errors make very uncertain to constrain the final number of experienced
mergers.

At present, there are a few cosmological simulations where the size growth of
the massive galaxies is explained by the accretion of minor satellites
\citep[see][]{Naab2009, Oser2011}. It would be straightforward to compare our
findings with those cosmological simulations of galaxy formation and check
whether the fractions that we find are recovered in such theoretical analysis.
If this was the case, the support to the minor merging mechanism as the main
responsible for the size evolution of the massive galaxies will be greatly
enhanced.

\section*{Acknowledgments}

We thank the anonymous referee for a careful and constructive reading of the
manuscript that help us to improve the quality of the paper. Authors are
grateful to Lulu Liu for providing us with their measurements of the fraction
of galaxies with satellites obtained from the SDSS, used here as a local
($z=0$) comparison. We thank Juan Betancort for his valuable input on several
aspects of the statistical analysis of this paper. We are devoted to Sergio
Pascual for his very useful help in programming questions. We would like also
to acknowledge fruitful discussions with Javier Cenarro, Luis D\'{\i}az, Rosa
Dom\'{\i}nguez, Cesar Gonz\'alez, Carlos L\'opez-San Juan, Jos\'e O\~norbe,
Thorsten Naab and Vicent Quilis. IT is a Ram\'on y Cajal Fellow of the Spanish
Ministry of Science and Innovation. This work has been supported by the
``Programa Nacional de Astronom\'{\i}a y Astrof\'{\i}sica'' of the Spanish
Ministry of Science and Innovation under grant AYA2010-21322-C03-02. PGP and GB
acknowledge support from the Spanish Programa Nacional de Astronomía y
Astrofísica under grants AYA2009-10368 and AYA2009-07723-E.  This work has made
use of the Rainbow Cosmological Surveys Database, which is operated by the
Universidad Complutense de Madrid (UCM).

\bibliography{E_Marmol_Queralto_revised}
\bibliographystyle{mn2e}

\end{document}